%% file: main_IEEE.tex
\documentclass{article}
\usepackage{arxiv}
\usepackage[utf8]{inputenc} 
\usepackage[T1]{fontenc}    
\usepackage{url}            
\usepackage{booktabs}       
\usepackage{amsfonts}       
\usepackage{amsmath}
\usepackage{amssymb}
\usepackage{siunitx}
\usepackage{nicefrac}       
\usepackage{microtype}      
\usepackage{lipsum}		
\usepackage{doi}
\usepackage[numbers]{natbib}
\usepackage[normalem]{ulem}
\usepackage{graphicx}
\usepackage{mathtools}
\usepackage{xcolor} 
\usepackage{wasysym}
\usepackage{cuted}
\usepackage{hyphenat}
\usepackage{makecell}
\usepackage{tablefootnote}
\usepackage[flushleft]{threeparttable}

\usepackage{bookmark}

\input{macros}

\title{2D hydrodynamic simulation of TeraFETs beyond the gradual-channel approximation for transient, large-signal or ultrahigh-frequency simulations}

\renewcommand{\shorttitle}{2D hydrodynamic simulation of TeraFETs beyond the gradual-channel approximation for transient, large-signal or ultrahigh-frequency simulations}

\author{Florian Ludwig\\
Physikalisches Institut\\
Johann Wolfgang Goethe-Universität\\
DE-60438 Frankfurt am Main, Germany \\
\texttt{ludwig@physik.uni-frankfurt.de} \\
   \And
Hartmut G. Roskos \\
Physikalisches Institut\\
Johann Wolfgang Goethe-Universität\\
DE-60438 Frankfurt am Main, Germany\\
\texttt{roskos@physik.uni-frankfurt.de} \\
\And
Raul Borsche \\
Department of Mathematics\\
RPTU Kaiserslautern-Landau\\
DE-67663 Kaiserslautern, Germany \\
\texttt{borsche@rptu.de}
}

\hypersetup{
pdftitle={2D hydrodynamic simulation of TeraFETs beyond the gradual-channel approximation for transient, large-signal or ultrahigh-frequency simulations},
pdfsubject={physics.app-ph, physics.cond-mat},
pdfauthor={Florian Ludwig, Hartmut G. Roskos, Raul Borsche},
pdfkeywords={TCAD, hydrodynamic, modeling, HLLC, Riemann solver, Si, CMOS, MOSFET, plasma wave, terahertz detection},
}

\begin{document}
\begin{titlepage}
\maketitle
\vspace{-1em}
\begin{abstract}
In the past decade, detection of THz radiation by plasma-wave-assisted frequency mixing in antenna-coupled field-effect transistors (TeraFETs) -- implemented in various semiconductor material systems (Si CMOS, GaN/AlGaN, GaAs/AlGaAs, graphene, etc.) -- has matured and led to a practically applied detector technology. This has been supported by the development of powerful device simulation tools which take into account relevant collective carrier dynamics and mixing processes in various approximations. These tools mostly model carrier transport in 1D and they are usually geared towards continuous-wave illumination of the device and small-signal response. Depending on their implementation, it may not be possible readily to simulate large-signal and pulsed operation. Another approximation which may lead to unsatisfactory results is the 1D restriction to calculate only the longitudinal electric field components. Especially at the edges of the gate electrode, solving of the 2D Poisson equation promises better results. This contribution introduces a stable way to solve the 2D Poisson equation self-consistently with the hydrodynamic transport equations including the numerically challenging convection term. We employ a well-balanced approximate Harten-Lax-van-Leer-Contact Riemann solver. The approach is well suited for a future treatment of transient and large-signal cases. The 2D treatment also generically extends the model beyond the gradual-channel approximation and allows to calculate the FET's response at high THz frequencies where the gate-to-channel potential acquires a non-local character. Model calculations are performed for the exemplary case of a 65-nm Si CMOS TeraFET in the isothermal approximation. 
\end{abstract}

\end{titlepage}

\section{Introduction}
Three decades ago, Dyakonov and Shur put forth the idea to use field-effect transistors (FETs) as resonators for plasma waves \cite{Dyakonov1993}. In their pioneering theoretical work they predicted that a steady current flow can become unstable against generation of gated quasi-two-dimensional (quasi-2D) plasma waves (plasmons) -- this phenomenon being known as Dyakonov-Shur instability \cite{Crowne1997,Mendl2018,Kargar2018} and predicted to be a possible source of high-frequency electromagnetic radiation.\footnote{\textcolor{black}{Although the presented tool can simulate these applications using the HDM, moment-based models may be inaccurate in the quasi-ballistic regime, where plasma-wave instabilities have been proposed. Monte Carlo solutions of the Boltzmann Transport Equation (BTE), incorporating accurate boundary conditions, suggest the absence of plasma resonances and question the Dyakonov-Shur instability. Thus, Monte Carlo simulations should be pursued in this regime \cite{noei_numerical_2020}.}} 
In a follow-up work by the same authors, it was shown that the nonlinear mixing properties of the gated 2D plasmons can be used to detect and frequency-mix THz radiation in the FET channel \cite{Dyakonov1996}. The mathematical basis for their predications was 
a small-signal harmonic analysis of a reduced 1D hydrodynamic transport model (HDM), which neglected carrier diffusion, carrier heating and the embedding of the gated channel in a real circuit environment \cite{Boppel2012}. 
While no unambiguous experimental evidence for the Dyakonov-Shur instability has been found until now \cite{Jungemann2022}, 
antenna-coupled FETs (TeraFETs) are successfully in use as sensitive detectors and mixers \cite{Knap2004, Lisauskas2009, Lisauskas2013mix, Glaab2010, BoppelCMOSDetector, Roadmap2021, yuan2023a, wiecha_antenna-coupled_2021}. TeraFETs are implemented in various material systems and device technologies 
such as Si CMOS \cite{Boppel2012, hillger_terahertz_2019, Ikamas2018},  III-V semiconductor-based high electron mobility transistors (HEMTs)\cite{Hou2017, Regensburger2018_CW, Bauer2019, Regensburger2024} and graphene FETs \cite{Zak2014, Generalov2017, Viti2020, Generalov2024}. 
TeraFETs are implemented with narrow-band or broadband antennae, have a detection speed in the sub-ns range \cite{Viti2021} and exhibit high sensitivity with values of the optical noise equivalent power (NEP) as low as 20~pW/$\surd$Hz \cite{Mateos2020, Generalov2024} at room temperature. 
That gated plasmons can be excited in TeraFETs by THz radiation has been confirmed by both transport measurements \cite{Drexler2012, Bandurin2018a, Delgado-Notario2024} and s-SNOM (scattering-type scanning near-field optical microscopy)  experiments \cite{Soltani2020}. 

The response of TeraFETs to continuous-wave THz radiation in the small-signal limit can be simulated on the basis of a 1D HDM, often in the form of an equivalent waveguide representation which can be embedded into a circuit design tool \cite{Ludwig2019, Ludwig2024_Si_ADS}. The small-signal THz transport processes can also be captured empirically, e.g. by means of a Volterra series approach \cite{Stake2016}. Quantitative agreement with the measured responsivity and NEP values is achieved. However, the models are not suitable to treat transient phenomena upon pulsed THz excitation, to determine the large-signal response, or to correctly describe the detector response at very high THz frequencies (above 10~THz). \textcolor{black}{For these purposes, current simulation methods primarily rely on Monte Carlo approaches \cite{Millithaler09, Millithaler11, Mateos12}, which require a large number of iterations to reduce inherent noise to acceptable levels, thus increasing computational cost \cite{oberman2011monte, dimarco2018uncertainty}. In contrast, grid-based models, such as the HDM, offer significant advantages, particularly in higher dimensions where capturing small-scale structures and intricate flow features is critical. To extend the HDM to these scenarios, we will thoroughly investigate the challenges that emerge when most common approximations are removed.}

The HDM consists of a set of equations associated with conservation laws. These equations can be derived from the Boltzmann transport equation using the method of moments \cite{Blotekjaer1970,grasser2003}. The two most commonly applied moments-based transport models, the Drift-Diffusion (DD) and the Energy-Balance (EB) model, incorporate two and three moments of the charge carriers' distribution functions, respectively. However, neither model includes the spatial derivative of the charge carriers' drift (kinetic) energy, which becomes significant at THz frequencies and in the presence of strong electric field gradients, leading to a convective response in the carrier ensemble. The numerical treatment of this convection is challenging and may fail to converge. For this reason, implementations of the models in commercially available computer-aided design software (e.g. from the companies Synopsys or Silvaco) neglect the convection term (unlike 1D treatments such as those in \cite{Gutin2012, Rudin2014, Ludwig2024_Si_ADS}). 
This approximation is often referred to as the diffusion approximation \cite{grasser2003}.

Mathematically, the resulting equation systems form a parabolic system of partial differential equations (PDEs). They can be solved numerically by means of well-known Scharfetter-Gummel (SG) discretization schemes \cite{Scharfetter1969,Selberherr1984}. DD and EB models in the diffusion approximation have become a backbone for the simulation of device physics of modern transistor technology. While they also allow for the simulation of TeraFETs at low THz frequencies (e.g., 0.3~THz, simulated using the Sentaurus model of Synopsys \cite{Liu2019}), it is expected that these simulations will deviate from experimental results as the excitation frequency is increased and plasma waves become important \cite{grasser2003,Ludwig2019,Linn2020,Ludwig2024_Si_ADS}. 

Mathematically, the inclusion of the convection term transforms the equation set into hyperbolic PDEs, where, contrary to the parabolic PDEs, the solutions of the equation system due to external disturbance are not instantaneous (infinite velocity) but travel with a finite characteristic velocity through the simulation domain. Under these conditions, the SG scheme fails. As pointed out in \cite{Linn2020}, an additional numerical problem, which arises specifically upon simulation of semiconductor devices with large built-in electric fields, e.g. arising from large doping gradients or large applied bias voltage, is that the computation cannot be stabilized with standard numerical schemes available in the field of computational fluid dynamics (CFD). For example, the first-order accurate upwind scheme or the second-order accurate Richtmyer Two-Step Lax-Wendroff method \cite{Leveque2007}, which have recently been applied  successfully to simulations of THz plasmonics \cite{Bhardwaj2018, bhardwaj_electronicelectromagnetic_2019,cosme_tethys_2023}, may lead to unphysical negative carrier densities  if not adapted accordingly \cite{kurganov_second-order_2007,kurganov_finite-volume_2018}. 
To tackle this problem, a splitting approach for a 1D hyperbolic DD model (extended DD or isothermal hydrodynamic model) was introduced, where the equations are split into a stationary and a dynamic part in order to achieve well-balanced behavior \cite{Linn2020}. Such a scheme is of particular interest for the numerical simulation of electronic THz components as it enables to resolve small perturbations around a stationary solution. 
The numerical schemes 
mentioned above as being used for simulations of THz plasmonics \cite{Bhardwaj2018, bhardwaj_electronicelectromagnetic_2019,cosme_tethys_2023} are not of the well-balanced kind. As pointed out in \cite{kurganov_finite-volume_2018}, well-balanced schemes are especially advantageous if coarse computational grids are applied to retrieve periodic steady-state solutions. In these situations, the magnitude of the truncation error of a non-well-balanced scheme may be larger than the magnitude of the waves to be modeled.

\textcolor{black}{In \cite{Linn2020}, the dynamic problem was solved by using either a piecewise constant reconstruction or a CWENOZ3 (third-order centered weighted essentially non-oscillatory) scheme \cite{Cravero2018,Cravero2019} together with an exact Riemann solver. However, this approach turned out to be numerically unstable and not positivity-preserving for large source terms. In this work, we also employ piecewise constant reconstruction, but use a well-balanced Harten-Lax-van-Leer-Contact (HLLC) Riemann solver \cite{Toro1994,Toro2009,Audusse2015} to simulate plasma waves. In contrast to the splitting approach in \cite{Linn2020}, our approximate Riemann solver is inherently well-balanced, eliminating the need for a computationally costly high-order stiff ordinary differential equation (ODE) solver to achieve well-balancedness in the stationary solution to nearly machine precision. The underlying transport model is the same as that presented in \cite{Linn2020} - an extended DD model (isothermal hydrodynamic transport model) --, but solved in 2D. As usual for these models, the electron and hole distributions are always assumed to be thermal, with the carrier temperatures equal to the lattice temperature $T_L$. Notably, the isothermal approximation may become unreliable for nanoscale devices particularly under large-signal (strong field) excitations, where carrier heating cannot be ignored. However, in \cite{Ludwig2024_Si_ADS} 1D hydrodynamic transport simulations of 65-nm Si CMOS TeraFETs relying on the isothermal approximation have shown to deliver quantitative agreement with measurements when considering small-signal excitations. An implementation of the full hydrodynamic transport model to correctly address carrier transport under large-signal conditions is subject to future work.}
In contrast to the 1D model used by Dyakonov and Shur in the 1990s, diffusive current contributions are taken into account. We find that the well-balanced HLLC Riemann solver is numerically stable and positivity-preserving even in case of enormous source terms. In our model calculations, we consider a 65-nm n-channel Si CMOS TeraFET detector embedded in a log-spiral antenna. Such devices, fabricated for us at the TSMC foundry, were characterized experimentally at room temperature (DC characteristics and THz detection), and were simulated in 1D both with a TSMC foundry model and our own ADS-HDM simulator \cite{Ludwig2024_Si_ADS}.

\section{Stationary transport model}
In order to obtain the stationary (DC) response of the THz detector, we solved the 2D Poisson equation \cite{Selberherr1984} together with the continuity equations for electrons and holes. The Poisson equation is given as 
\begin{align}
    \nabla \cdot (\epsilon\nabla \Psi) = -q(p-n+C_{net}) \, .
    \label{eq:Poisson}
\end{align}
Here, $\epsilon$ is the dielectric permittivity, $\Psi$ the quasi-stationary electric potential, $q$ the absolute value of the elementary charge,$n$ and $p$ the electron and hole density, respectively.
$C_{net} = N_D^+-N_A^-$ is the local net doping density\footnote{A standard incomplete-ionization model for silicon was taken into account \cite{Cole1989} to determine the densities $N_D^+$ and $N_A^-$ of the ionized impurities.}. The continuity equations are  \cite{Selberherr1984,HandbookOptoSim}
\begin{align}
    \nabla(j_n) = \nabla(-q\mu_n n \nabla\varphi_n) = q R \label{eq:DDe} \,,\\   
    \nabla(j_p) = \nabla(-q\mu_p p \nabla\varphi_p) = -q R  \,, \label{eq:DDh}
\end{align}
written in terms of the quasi-Fermi potentials (or electrochemical potentials) for electrons ($\varphi_n$) and holes ($\varphi_p$). Here, $\mu_n$ and $\mu_p$ represent the high-field mobilities determined by the Caughey-Thomas mobility model \cite{Caughey1967}. This set of non-linear PDEs is also known as the Van Roosbroeck system\cite{VanRoosbroeck1950}. For the calculation of the low-field mobilities $\mu_{n,0}$ and $\mu_{p,0}$ of the electrons and holes, we took scattering with phonons\cite{Lombardi1988}, impurities\cite{Masetti1983} as well as surface-phonon and surface-roughness scattering at the Si/SiO$_{2}$ interface into account\cite{Lombardi1988}. The recombination rate $R$ was modeled after the Shockley-Read-Hall model\cite{Shockley1952}. We explicitly chose the quasi-Fermi-potential formalism for the stationary solver, since we intend to model not only bulk silicon, but in the future also high-electron-mobility transistors (HEMTs) based on III-V compound semiconductors and graphene FETs with the modeling platform presented here (acronym: TeraCAD). The current equations Eqn.~(\ref{eq:DDe}) and (\ref{eq:DDh}) can be applied directly for heterostructure systems, whereas a carrier-density-based formalism needs an explicit modification due to the position-dependent eﬀective density of states and band edges \cite{AufderMaur2008}. 

The equations were discretized using a vertex-centered finite volume method and all equations presented above and in the following section were subject to appropriate scaling \cite{vasileska_computational_2017,Selberherr1984}. The coupled equations were solved with the Newton Successive Over-Relaxation method \cite{Selberherr1984} in combination with a line-search algorithm for the relaxation factor \cite{NumericalRecipes2007}. The carrier densities (e.g. $n_{i+1/2,j}$, $p_{i+1/2,j}$) at the cell interfaces were calculated based on the assumption that the electric ($\Psi$) and quasi-Fermi potentials ($\varphi_n$, $\varphi_p$) vary linearly between two adjacent cells \cite{AufderMaur2008}. For Boltzmann statistics, we obtain
\begin{align}
    n_{i+1/2,j} = N_c \cdot \exp{\left(\eta_{n;i+1/2,j}\right)} \,, \\  
    \eta_{n;i+1/2,j} = \frac{q\left(\Psi_{i+1/2,j}-\varphi_{n;i+1/2,j}\right)-E_{c}}{k_B T_{C,n}}\,.
\end{align}
\begin{figure}[!t]
\centering
\includegraphics[width=4.5in]{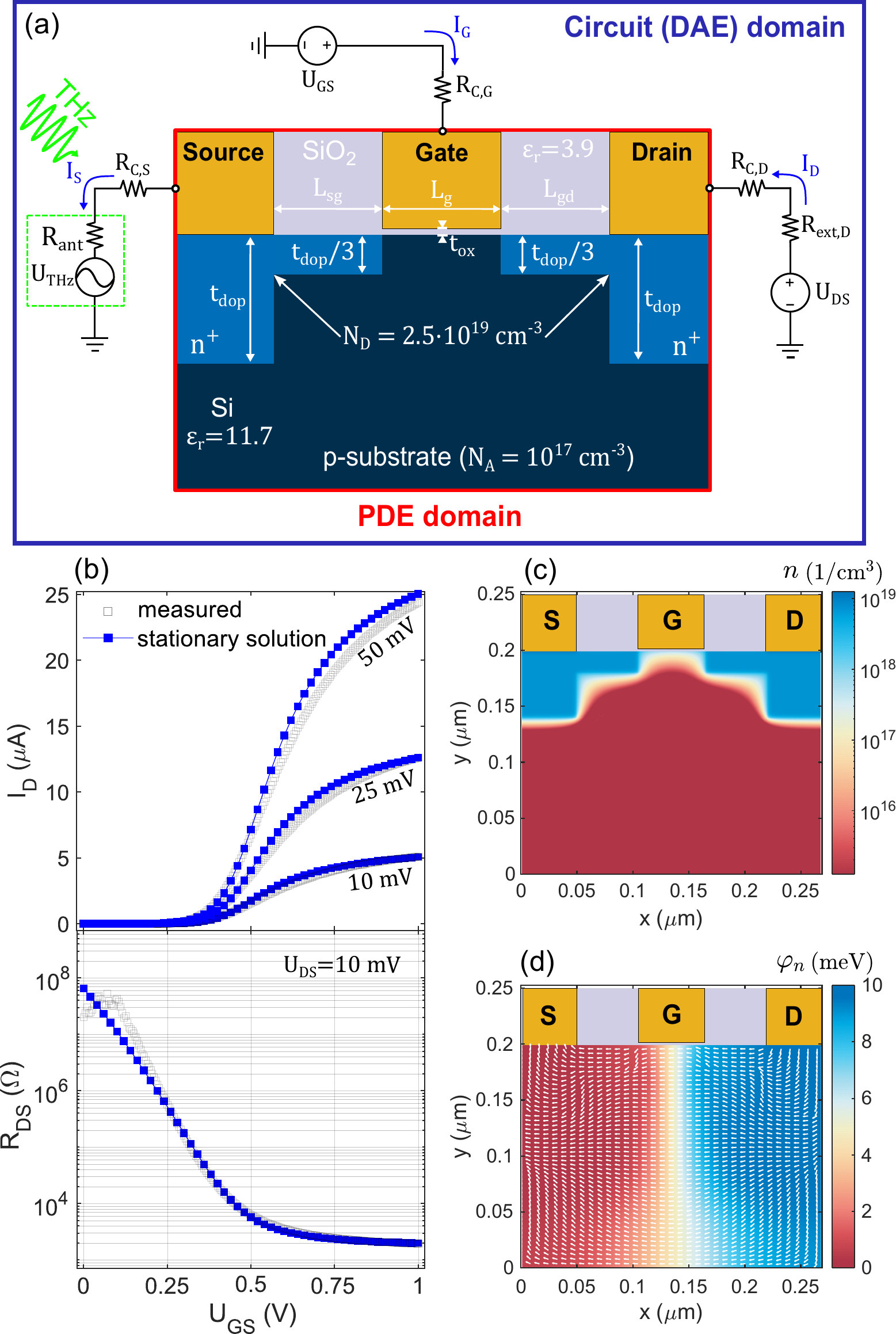}
\caption{(a) Schematic of the TeraCAD simulation tool: The blue box indicates the full simulation domain including the outer circuitry, here presented for source coupling conditions. The red box depicts the PDE domain, where the hydrodynamic transport equations are solved. The key geometric parameters used for the 65-nm MOSFET can be found in Table~\ref{tab:FETparameter}.\\ (b) Comparison of stationary (DC) TeraCAD simulation results with experimental characterization data for a 65-nm Si CMOS TeraFET. \textcolor{black}{Shown are the current $I_{D}$ at the drain terminal (top panel) for three different drain-to-source voltages $U_{DS}$ and the drain-to-source resistance $R_{DS} = U_{DS}/I_{D}$ (bottom panel) for $U_{DS}=10$~mV as a function of the gate voltage $U_{GS}$.} \\ (c,d) 2D distributions of the stationary electron density $n$ and the electron quasi-Fermi potential $\varphi_{n}$ for $U_{DS}=10$~mV and $U_{GS}=0.5$~V. In (d), the overlaid white arrows depict the local flow direction of the electron current.}
\label{fig:Schematic+DC}
\end{figure}

\begin{table}[!t]\centering
\caption{Key geometric simulation parameters for the 65-nm Si MOSFET. The total PDE domain size is 0.27~$\mu$m $\times$ 0.253~$\mu$m (length $\times$ height). }
\begin{threeparttable}
\begin{tabular}{clc}\hline
Parameter & Description & Value \\
\hline \hline
$W$ & channel width & 0.7 $\mu$m  \\ 
 \hline
$L_g$ & gate length  & 60~nm \\ 
\hline
$L_{sg}$ and $L_{gd}$ & \makecell[l]{gate-to-contact \\ metallization distance}  & 55~nm \\ 
\hline
$t_{ox}$ & \makecell[l]{gate-to-channel \\ oxide thickness\tnote{1}}
& 3~nm \\ 
\hline 
$t_{dop}$ & doping depth & 60 nm \\ 
\hline\end{tabular}\label{tab:FETparameter}%
\begin{tablenotes}
\item[1] \vspace{2mm} Here, $t_{ox}$ represents the effective insulator thickness ($t_{ox} = t_{phys} + t_{qc}$), which is the sum of the physical insulator thickness ($t_{phys} = 2.6~\rm{nm}$) and a quantum correction of $t_{gc} \approx 0.4~\rm{nm}$ due to the quantum-confinement effect of electrons at the Si/SiO\textsubscript{2} interface in nanoscale MOSFETs \cite{Saad2010}. $d_{phys} = 2.6~\rm{nm}$ is taken from the information sheet of TSMC for the 65-nm CMOS foundry process.

\end{tablenotes}
\end{threeparttable}
\end{table}%
Here, $N_c=2(2 \pi k_B T_{C,n} m^*_{n,DOS}/h^2)^{3/2}$ is the effective density of states in the conduction band, with $k_B$ being the Boltzmann constant, $T_{C,n}$ the carrier temperature of the electrons, and $m^*_{n,DOS}=1.08 \, m_0$ the density-of-states effective mass of electrons in Si 
($m_0$: mass of free electrons) \cite{green_intrinsic_1990}. As already mentioned, we assumed 
$T_{C,n} = T_{C,p} = T_L = 294\,$K (room-temperature). $E_c=E_{c,0}-E_{F,i}$ represents the effective conduction band-edge energy, where $E_{c,0}$ is the conduction band-edge for the pure material and $E_{F,i}$ the intrinsic Fermi level. Note that here, $N_c$ and $E_c$ were assumed to be constant throughout the semiconductor. For heterostructure systems, these quantities can vary in space and need to be interpolated onto the cell interfaces as in case of $\Psi_{i+1/2,j}$ and $\varphi_{n;i+1/2,j}$. 

\Figrefa{fig:Schematic+DC} presents the schematic of the simulated device. The geometric parameters are given in Table~\ref{tab:FETparameter}. As the actual net doping concentration and profile ($C_{net} (x,y)$) were not known (they represent undisclosed information of the TSMC foundry), we determined $C_{net} (x,y)$ such that the simulated $R_{DS}(U_{GS})$ curve of the stationary model is in fair agreement with the measured one, 
as shown in \Figrefb{fig:Schematic+DC}. For all simulations carried out in this work we used an uniform grid with $\Delta x= 1\,$nm and $\Delta y= 0.5\,$nm spacing. For the stationary simulations, we assumed Dirichlet boundary conditions for $\Psi$, $\varphi_n$ and $\varphi_p$ at the semiconductor-metal interfaces (ideal ohmic contacts) \cite{Selberherr1984}, thereby fixing the electron density $n$ at the contacts (visualized in \Figrefc{fig:Schematic+DC}). 
We treated the gate electrode (a poly-Si gate stack \cite{Fung2004,Yang2011}) as an equipotential surface 
with an effective work function of $4.53\,$eV (as for a  poly-Si/TiN/SiO2 stack  \cite{Singanamalla2007,Kadoshima2009,Robertson2009}). In order to close the unbounded PDE domain (red box in \Figrefa{fig:Schematic+DC}) and to limit the computational load, we introduced artificial boundary conditions of von-Neumann type \cite{Selberherr1984} for $\Psi$ at the PDE domain edges and for $\varphi_n$ and $\varphi_p$ at the semiconductor domain edges (including the internal semiconductor-insulator (Si/SiO\textsubscript{2}) interface).\footnote{
The artificial boundary of $\Psi$ 
was chosen such that an increase of 
the PDE domain size 
did not significantly change the simulated resistance curve anymore. 
The von-Neumann boundary conditions imply that outward-orientated fluxes are prohibited and the PDE domain remains self-contained.} \Figrefd{fig:Schematic+DC}) displays the spatial dependence of the simulated electron quasi-Fermi potential for $U_{DS}=10\,$mV. The local direction of the electron current is indicated by white arrows. 

The PDE domain was coupled to an external driver circuit (blue box in \Figrefa{fig:Schematic+DC}), similar to that used in \cite{Jungemann2016,Linn2020}, which represents a differential algebraic equation (DAE) that can be derived from Kirchhoff's current and voltage laws. \textcolor{black}{For the DC simulations of the current-voltage curve shown in the top panel of \Figrefb{fig:Schematic+DC}, $U_{DS}$ was fixed at three different values while $U_{GS}$ was varied. We achieve quantitative agreement between measurements and simulations within the available $U_{DS}$ range.}\footnote{\textcolor{black}{A complete validation of the stationary solver would require further DC measurements and simulations at higher $U_{DS}$ values. However, the primary focus of this work is on the numerical treatment and implementation of the HDM.}} 
For the THz simulations, 
we coupled the THz wave to the source electrode 
(as indicated by the green box in \Figrefa{fig:Schematic+DC}). The currents (including the displacement currents) at the source, drain and gate electrodes were calculated using the Ramo-Shockley theorem \cite{Ramo1939,Kim1991}. For all simulations (DC and THz), we assumed a constant and real-valued antenna impedance of $R_{ant} = 100~\Omega$, an external parasitic resistance $R_{ext,D} = 200~\Omega$ (corresponding to the approximate residual resistance of a protection FET in the real device), and a contact resistivity $\rho_c = 5~\rm{m\Omega\, cm}$, which results in a contact resistance of $R_{C,S}=R_{C,D}=R_{C,G} = 71.4~\Omega$ for the given geometric parameters (see Table~\ref{tab:FETparameter}).

\section{Hydrodynamic transport model}
For the transient hydrodynamic simulations of n-channel Si MOSFETs, we considered only the dynamic response of the electrons and fixed the hole density in time to its respective stationary solution obtained for each gate voltage.\footnote{\textcolor{black}{The influence of hole carrier dynamics on the detector response can be neglected, as the hole density is much lower than the electron density at the detector's optimal gate bias conditions. The deviation of the simulated DC current obtained for stationary model with and without holes is less than 30\% at $U_{GS} = 0.5 V$ and approximately 4\% at $U_{GS} = 1 V$.}} The 2D isothermal HDM for electrons is given in conservation form by
\begin{align}
    \partial_t \boldsymbol{u} - \partial_x \boldsymbol{F}(\boldsymbol{u}) - \partial_y \boldsymbol{G}(\boldsymbol{u}) = \boldsymbol{S}(\boldsymbol{u}) \,,
    \label{HDM}
\end{align}
with
\begin{align*}
\boldsymbol{u} = \begin{pmatrix}n\\[1.5mm]j_x\\[1.5mm]j_y\end{pmatrix},\, \boldsymbol{S}(\boldsymbol{u}) = \begin{pmatrix}0\\[1.5mm]\left[\frac{qE_x}{m^*_n}\right] n - \frac{j_x}{\tau_{p,n}}\\[1.5mm]  \left[\frac{qE_y}{m^*_n}\right] n - \frac{j_y}{\tau_{p,n}}\end{pmatrix},\, \\
\boldsymbol{F}(\boldsymbol{u}) = \begin{pmatrix}j_x\\[1.5mm]\left[\frac{k_B T_L}{m^*_n}\right] n + \frac{j_x^2}{n}\\[1.5mm]\left(\frac{j_yj_x}{n}\right)\end{pmatrix},\, \boldsymbol{G}(\boldsymbol{u}) = \begin{pmatrix}j_y\\[1.5mm]\left(\frac{j_xj_y}{n}\right)\\[1.5mm]\left[\frac{k_B T_L}{m^*_n}\right] n + \frac{j_y^2}{n} \end{pmatrix}\,.
\end{align*}
Here, $E_x=- \partial_x \Psi$ and $E_y = -\partial_y \Psi$ represent the electric fields, and $j_x=-qnv_x$ and $j_y=-qnv_y$ the current densities, with $v_x$ and $v_y$ being the velocity components in $x$- and $y$-direction, respectively. They 
were determined from the eigenvalues of the differential operators $\boldsymbol{F}(\boldsymbol{u})$ and $\boldsymbol{G}(\boldsymbol{u})$. They are defined by
\begin{align}
\begin{split}
&\lambda_{F,1;3} =  v_x \mp c, \quad \lambda_{F,2} =  v_x, \\
&\lambda_{G,1;3} =  v_y \mp c, \quad \lambda_{G,2} =  v_y,
\end{split}
\label{eq:wavespeeds}
\end{align}
where $c=\sqrt{k_B T_L/m^*_n} \sim10^5\,$m/s is the thermal (sound) velocity. The $\lambda_{F,1;2;3}$ and $\lambda_{G,1;2;3}$ are to be determined by the differential operators calculated from the unknown quantities $\boldsymbol{u}$ obtained in the previous time step.
For the scattering terms $j_x/\tau_{p,n}$ and $j_y/\tau_{p,n}$ the electron momentum relaxation time $\tau_{p,n}$ was determined from the high-field electron mobility via $\tau_{p,n} = \mu_n \cdot m^*_n/q $, with the effective mass $m^*_{n}=0. 258\,m_0$ and $\mu_n$ are defined by the scattering mechanisms discussed in the previous section. 

To solve the multi-dimensional PDE system, we used the dimensional splitting approach (or fractional-step method) \cite{Yanenko1971} with additional source-term splitting \cite{Toro2009}, 
applied for each direction separately. The scattering terms (e.g. $-j_x/\tau_{p,n}$ of the $x$-split) in $\boldsymbol{S}(\boldsymbol{u})$ was accounted for through an implicit Euler method \cite{Leveque2007}. The drift terms (e.g. $qE_x n/m_n^* $ of the $x$-split) were incorporated in the well-balanced flux treatment as discussed below. 

\begin{figure}[!t]
\centering
\includegraphics[width=4.5in]{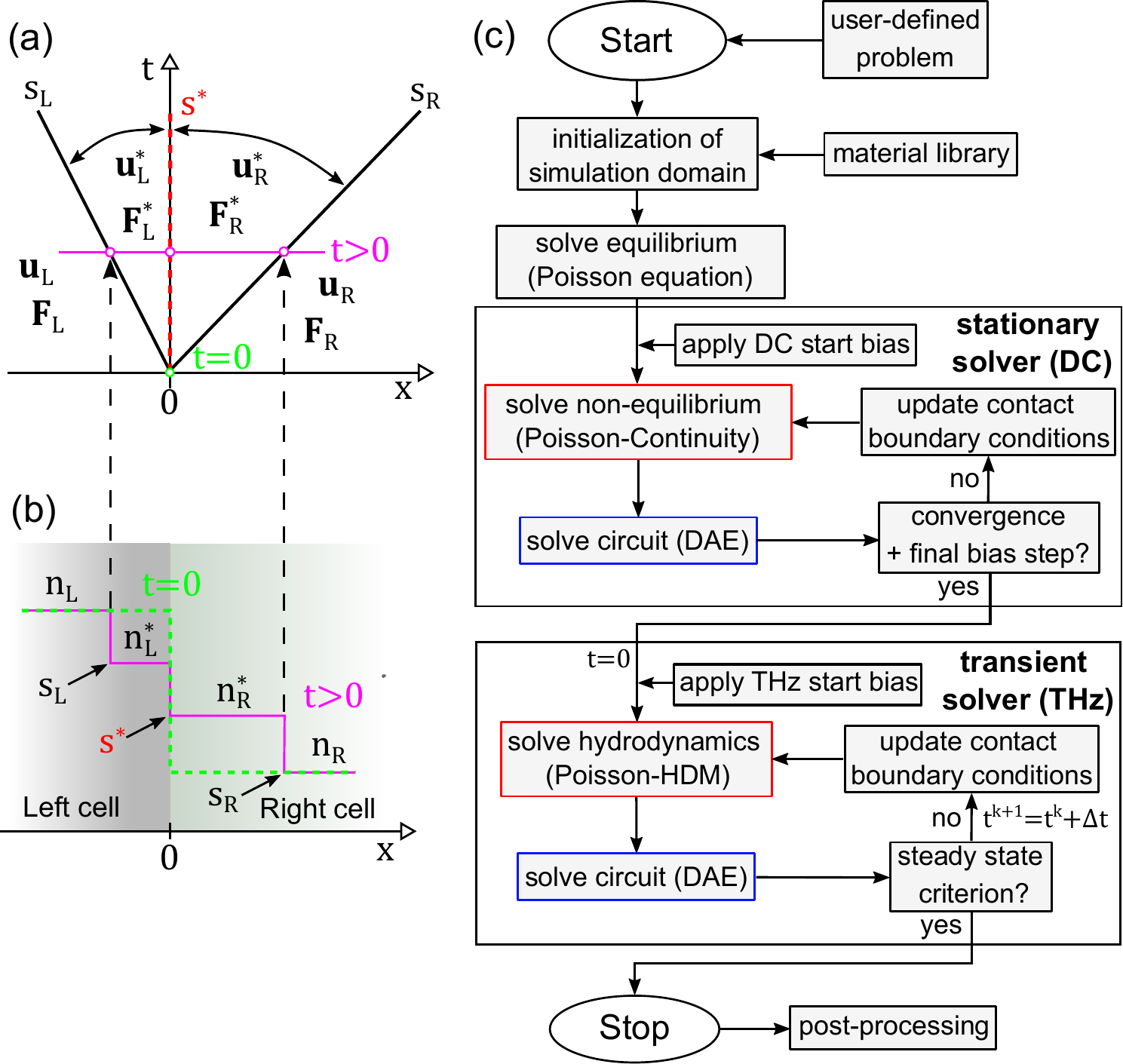}
\caption{(a) Time evolution of the three-wave Riemann problem for the $x$-split step, which is considered for the well-balanced HLLC  approximate Riemann solver in this work. (b) Sketch of a Riemann problem (a single discontinuity at intital time $t=0$, marked as green dashed line), between two adjacent cells containing the electron densities $n_L$ and $n_R$. For the three-wave HLLC scheme, the intermediate states $n^*_L$ and $n^*_R$ are introduced. To facilitate comprehension, we use two dashed arrows to link the corresponding left ($s_L$) and right ($s_R$) shock waves at a fixed point in time $t>0$ (magenta solid line) between panels (a) and (b). Under well-balanced conditions, one assumes a static contact wave with $s^*=0$. \\ (c) A simplified flow chart for the TeraCAD simulations.}
\label{fig:Riemann+FlowChart}
\end{figure}

We tested two well-balanced schemes for the plasma wave simulations: 
The f-wave path-integral method \cite{LeVeque2011} and the HLLC method \cite{Toro1994,Toro2009,Audusse2015}.
Both 
are of upwind type (Riemann-problem-based), where information of the wave propagation is used to construct the numerical fluxes \cite{Toro2009}. The HLLC method was found to be numerical stable and positivity-preserving for $n$ when applying large external fields together with strong doping gradients.\footnote{The f-wave path integral method was found to be stable only for the subsonic case. We emphasize here that both applied methods should be suitable for large-signal simulations of plasma waves, and the observed instability of the f-wave path method implemented in TeraCAD could not be investigated in course of the presented work.}
In \Figrefa{fig:Riemann+FlowChart}, the Riemann problem for the HLLC method is depicted considering two adjacent cells (L = Left cell, R = Right cell). A Riemann problem is a special initial value problem (IVP) that contains a single discontinuity in the state variables $\boldsymbol{u}$ at the initial time $t=0$ (shown here as an example for the electron density as green dashed line in \Figrefb{fig:Riemann+FlowChart}) that is considered together with the hyperbolic PDE. The solution to the Riemann problem results in shock waves that travel with a characteristic velocity in ether positive ($s_R$) or negative ($s_L$) direction along with intermediate states that exists between them. For the HLLC method two intermediate states, $\boldsymbol{u}^*_L=(n^*_L,j^*_{x,L},j^*_{y,L})^T$ and $\boldsymbol{u}^*_R=(n^*_R,j^*_{x,R},j^*_{y,R})^T$, of the unknown vector $\boldsymbol{u}$ are considered. These states are separated by the so-called contact wave ($s^*$). 

Before determining the HLLC fluxes $\boldsymbol{F}^*_L$ and $\boldsymbol{F}^*_R$ for the intermediate states, one seeks for a well-balanced condition of the isothermal HDM, that is, an exact equilibrium of carrier diffusion and carrier drift motion determined from the stationary solution ($\partial_t \boldsymbol{u} = 0$, $\boldsymbol{j} = (j_x,j_y)^T = 0$) of the HDM. For the $x$-split, we have
\begin{align}
   \frac{1}{n} \partial_x (n) = \left[\frac{1}{U_T}\right]  \cdot \partial_x (\Psi) \,
    \label{WBcond}
\end{align}
where $U_T=k_B T_L/q$ represents the thermal voltage. Integrating between the adjacent cells centered at positions $x_L$ and $x_R$, one obtains
\begin{align}
    \ln{\left(\frac{n_R}{n_L}\right)} = \frac{\Psi_R-\Psi_L}{U_T} \,,
    \label{WBeq}
\end{align}
where $n(x=x_R) = n_R$ and $\Psi(x=x_R) = \Psi_R$, and  $n_L$, $\Psi_L$ accordingly. Eqn.~(\ref{WBeq}) represents our well-balance condition between two adjacent cells. Taking Eqn.~(\ref{WBeq}) into account and assuming a contact wave with speed $s^*=0$ between the intermediate states, we can determine the intermediate carrier densities ${n}^*_L$ and ${n}^*_R$ (as shown in \Figrefb{fig:Riemann+FlowChart}) from
\begin{align}
\begin{split}
    n_L^* & = \frac{n_{HLL}\cdot(s_R-s_L)}{s_R \cdot \exp{\left((\Psi_R-\Psi_L)/U_T\right)}-s_L} \,, \\
    n_R^* & = \frac{n_{HLL}\cdot(s_R-s_L)}{s_R-s_L\cdot \exp{\left(-(\Psi_R-\Psi_L)/U_T\right)}} \,,
\end{split}
\end{align}
and the intermediate current density ($j_L^*=j_R^*=j^*$) from
\begin{align}
\begin{split}
    j^* = j_{HLL} - \frac{\frac{q}{m_n^*} n_{Roe} (\Psi_R - \Psi_L)}{s_R-s_L}\,.
\end{split}
\label{Eq6}
\end{align}
$s_L=\min{\left(\lambda_{L,1},\lambda_{R,1},0\right)}$ and $s_R=\max{\left(\lambda_{L,3},\lambda_{R,3},0\right)}$ represent the wave velocities in the left and right cell (see \Figref{fig:Riemann+FlowChart}(a) and (b)), which are calculated from the characteristic speeds (Eqn.~(\ref{eq:wavespeeds})) $\lambda_{F,1;2;3}$ ($x$-split) or $\lambda_{G,1;2;3}$ ($y$-split). The unknown HLL vector components $n_{HLL}$ and $j_{HLL}$ are given by \cite{Harten1983,Bouchut2005,Toro2009}
\begin{align}
\begin{split}
    \boldsymbol{u}_{HLL} = \frac{s_R \boldsymbol{u}_R -s_L \boldsymbol{u}_L + \boldsymbol{F}(\boldsymbol{u}_R) - \boldsymbol{F}(\boldsymbol{u}_L)}{s_R-s_L} \,.
\end{split}
\end{align}
Note for our well-balanced scheme, the source term component due to the electric field ($q/m_n^* n_{Roe} (\Psi_R - \Psi_L)$) is directly included in the calculation of the intermediate current density (see Eq.~\ref{Eq6}). 
To ensure well-balancedness in thermal equilibrium, we need to calculated $n_{Roe}$ in the source term component by the so-called \textit{Roe average} \cite{Toro2009}. For the isothermal HDM with the well-balance property of Eqn.~(\ref{WBeq}), we obtain
 \begin{align}
	{n}_{\text{Roe}} = \frac{{n}_R-{n}_L}{\ln{\left(\frac{{n}_R}{{n}_L}\right)}} \, .
		\label{eq:n_roe}
\end{align} 
Finally, the HLLC fluxes can be obtained \cite{Toro2009} from
\begin{align}
\begin{split}
    \boldsymbol{F}^*_L = \boldsymbol{F}_L + s_L \cdot \left(\boldsymbol{u}_L^* -  \boldsymbol{u}_L\right) \,, \\
    \boldsymbol{F}^*_R = \boldsymbol{F}_R + s_R \cdot \left(\boldsymbol{u}_R^* -  \boldsymbol{u}_R\right) \,,
\end{split}
\end{align}
which are used to update the unknown vector $\boldsymbol{u}$ through a conservative update. For the $x$-split \cite{Toro2009,Audusse2015}:
\begin{align}
\begin{split}
    \boldsymbol{u}^{k+1}_i = \boldsymbol{u}^{k}_i  - \frac{\Delta t^k}{\Delta x} \left[\boldsymbol{F}_{L,i+1/2}^* -\boldsymbol{F}_{R,i-1/2}^* \right] \,.
\end{split}
\end{align}
For the $y$-split, the update is performed in an analogous way. Note that, for numerical stability, the time step $\Delta t^k$, where $k$ is the index of iteration, has to be estimated from the Courant-Friedrichs-Lewy (CFL) condition \cite{Toro2009}
\begin{align}
\begin{split}
    \Delta t^k = C_{cfl} \cdot \min\left(\frac{\Delta x}{\max{(|\lambda_{F,1;2;3}^k}|)},\frac{\Delta y}{\max{(|\lambda_{G,1;2;3}^k|)}}\right) \,,
\end{split}
\end{align}
where $C_{cfl}\in(0,1]$ is the CFL coefficient, which is set to $C_{cfl}=0.95$.

\section{Simulation of the detector responsivity}
\begin{figure}[!t]
\centering
\includegraphics[width=4.51in]{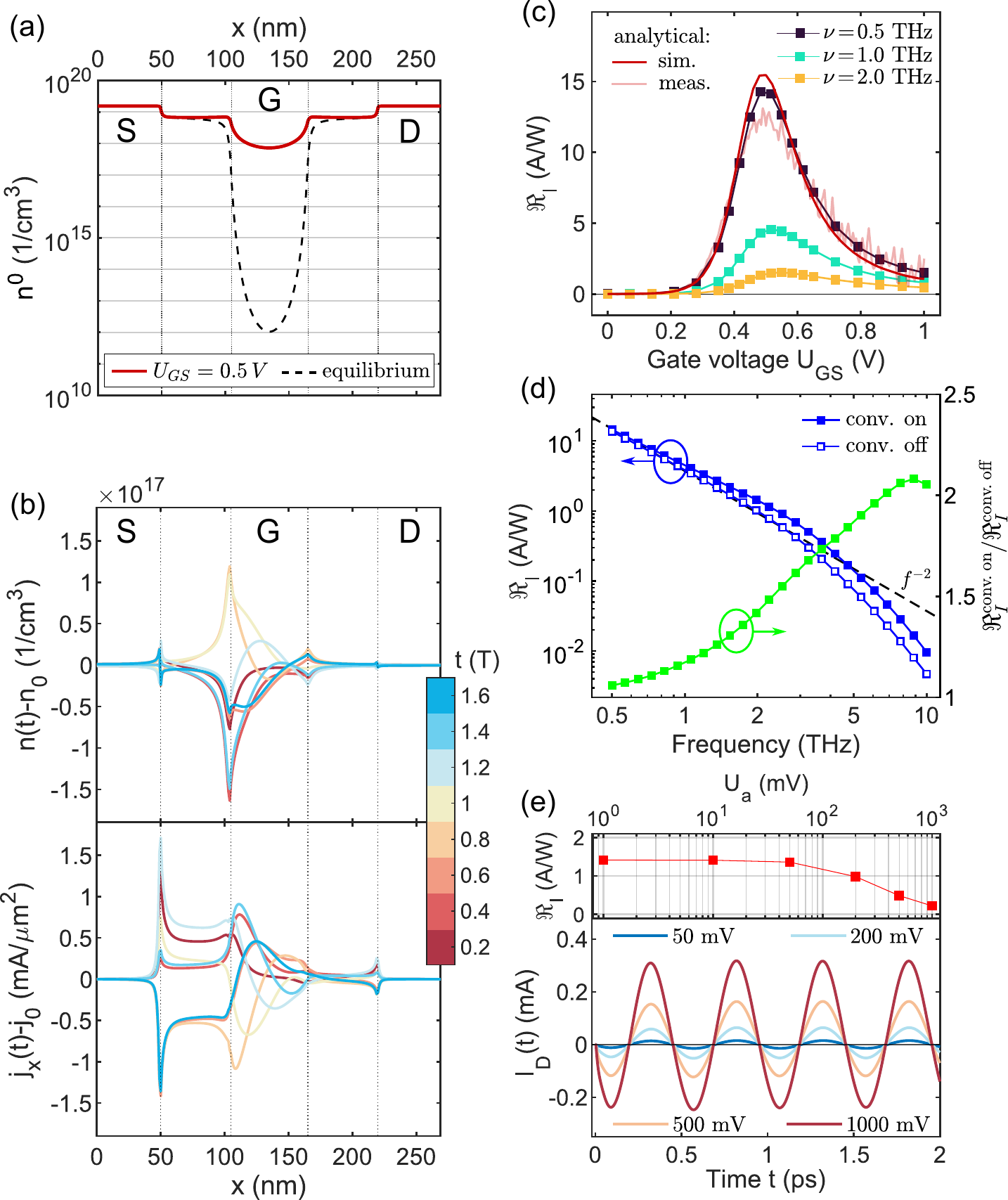}
\caption{(a) Stationary solver. Initial carrier density $n^0$ under equilibrium conditions ($U_{DS}=0.0\,$V) alongside the inversion layer of the MOSFET for $U_{GS}=0.5\,$V (red line). The thermal equilibrium (Poisson eq., $U_{GS}=0.0\,$V) is depicted as black dashed line. The different channel regions for Source contact (S), Gate (G) and Drain contact (D) are indicated by vertical dotted lines. \\ (b) Transient solver. Top panel: Oscillation of the carrier density $n(t)$ with regards to equilibrium $n(t=0)=n^0$ as a function of time (in units of $T$, color-coded). Bottom panel: Oscillation of the current density in x direction $j_x(t)$ with regards to equilibrium $j_{x}(t=0)=j_{x}^0$. For both panels $\nu = 2\,$ THz, $U_{GS}=0.5\,$V and $U_{a}=10~$mV. \\ \textcolor{black}{(c) Gate-voltage-dependent current responsivity $\Re_I$ for different frequencies under small-signal conditions ($U_{a}=1~$mV). The analytical approximations for resistive self-mixing detector response at $\nu = 0.5\,$ THz (calculated with Eq.~\ref{eq:Sako}) are depicted for the simulated (dark red line) and measured (bright red line) electrostatic transport characteristics obtained at $U_{DS}=10~$mV. \\ (d) Frequency-dependent current responsivity $\Re_I$ for a fixed gate voltage of $U_{GS}=0.5\,$V under small-signal conditions ($U_{a}=1~$mV). Simulations are performed with (full blue squares) and without (open blue squares) the convection term ($j^2/n$) to determine the importance of plasmonic effects. A black dashed line indicates a $f^{-2}$ roll off of the simulated current responsivity in the lower THz frequncy regime. The ratio of the simulated current responsivity for both cases ($\Re_I^{\rm{conv. on}}/\Re_I^{\rm{conv. off}}$) is depicted as green line (right vertical axis).} \\
(e) Time-dependent Drain terminal current $I_D(t)$ as a function different THz amplitudes $U_a$ for a sinusoidal excitation at $\nu = 2\,$ THz and $U_{GS}=0.5\,$V. The top inset depicts the periodic steady-state current responsivity as a function of $U_a$.}
\label{fig:Sim}
\end{figure}

Using the model described above, we simulated the detector response of a 65-nm Si CMOS TeraFET. In \Figrefc{fig:Riemann+FlowChart}, the flow chart of the TeraCAD calculations is presented. After initialization of the simulation domain, the stationary solver was applied for different gate voltages $U_{GS}$. Since TeraFETs are in most cases operated without drain-to-source bias, $U_{DS}=0~$V, they are by definition always in equilibrium before THz excitation. In \Figrefa{fig:Sim}, the resulting stationary carrier density $n^0$ in the channel of the MOSFET is plotted for $U_{GS}=0~$V and $U_{GS}=0.5~$V. Then, in the transient (HDM) solver, a sinusoidal excitation $U_{THz}(t)=U_a\cdot \sin(2\pi \nu \cdot t)$ was applied to the source electrode (see \Figrefa{fig:Schematic+DC}), where $\nu$ is the frequency and $U_a$ the amplitude of the THz voltage. The external excitation leads to a change in $\Psi$ (as a consequence of the different boundary conditions at the contacts) which was calculated from the dynamic Poisson equation \cite{Linn2020}
\begin{align}
    \nabla \cdot (\epsilon\nabla \Psi_D^k) = q(n_D^k) \, ,
\end{align}
where $n_D^k$ is the dynamic electron density $n_D^k(t) = n^k(t)-n^0$ and $\Psi_D^k$ the dynamic potential $\Psi_D^k(t)=\Psi^k(t)-\Psi^0$. $n^0$ (as shown in \Figrefa{fig:Sim}) and $\Psi^0$ correspond to stationary solutions of the PDEs. 
The PDE system was solved in sequence with the outer circuitry (see \Figrefa{fig:Schematic+DC}), which represents a stiff differential algebraic equation (DAE) system of the form
\begin{align}
    \boldsymbol{A}\cdot\boldsymbol{U} + \boldsymbol{B} \cdot \partial_t \boldsymbol{U} = \boldsymbol{b} 
\end{align}
derived from Kirchhoff's laws and solved by an appropriate ODE solver. Note that the ``mass'' matrix $\boldsymbol{B}$ contains all capacitance coefficients of the different metal electrodes, which are needed for the calculation of the displacement currents. The capacitances were determined from the Laplace equation (Eqn.~(\ref{eq:Poisson}) for zero space charge) \cite{Kim1991}. 

The periodic change of $\Psi$ 
leads to the development of plasma waves in the gated channel. To visualize these plasma waves, we plot in \Figrefb{fig:Sim} the evolution of $n$ (top panel) and $j_x$ (bottom panel) with respect to their spatially dependent equilibrium values 
at different time steps (in units of $T=1/\nu$). The simulations were performed  for $\nu = 2~$THz. 
One observes an overdamped plasma wave, which is expected, as the value of $\tau_{p,n}$ was found to be $\sim$60~fs in the gated channel region and $\sim$12~fs in the highly doped contact regions, leading into $\omega \tau_{p,n}<1$. \\
Next, we determined the intrinsic detector efficiency, 
expressed in terms of 
the current responsivity $\Re_I$, which can be obtained \cite{Jungemann2016,Linn2020} from
\begin{align}
    \Re_I &= \frac{\Delta I}{P_S^{THz}} \,,
\end{align}
where $\Delta I$ is the rectified (DC) current arising from the plasma-wave-assisted distributed resistive mixing \cite{Boppel2012, Oeje09} and $P_S^{THz}$ is the THz (AC) power, which is coupled from the source electrode into the FET. These quantities were obtained as 
\begin{align}
    \Delta I(t) &= \frac{1}{T} \cdot \int_{t-T}^t I_D(t) \rm{d}t \,, \\
    P_S^{THz} (t) &= \frac{1}{T} \cdot \int_{t-T}^t U_S(t) I_S(t) \rm{d}t \,,
\end{align}
where $I_D(t)$ and $I_S(t)$ are the drain and source currents, respectively, and $U_S(t)$ is the voltage at the source terminal. Note that the time dependence of the quantities $\Delta I(t), \; P_S^{THz} (t)$ arises for solutions, which have not yet reached the steady state.
As periodic steady-state criterion, we used in our simulations the relative deviation $\epsilon_{s}=\left(|\Delta I(t-T)- \Delta I(t)|\right)/|\Delta I(t)|$ and demanded 
that $\epsilon_{s}\leq \epsilon_{tol}=0.005$. 

In \Figrefc{fig:Sim}, the simulated current responsivity is presented as function of the gate voltage for $\nu = 0.5$, 1 and 2 THz. In order to verify that our 2D HDM simulations yield correct values of the responsivity, we compare it with the expected low-frequency response of the detector calculated by 
\begin{align}
\Delta I_{calc.} &= \frac{U_a^2}{4} \frac{1}{R_{DS}} \cdot \frac{\rm{\partial ln (I_{D})}}{\partial U_{GS}} \,.
\label{eq:Sako}
\end{align}
\textcolor{black}{This equation, derived in \cite{Sakowicz2011} through a small-signal analysis of reduced transport equations that account only for the drift motion of charge carriers - i.e., considering the resistive self-mixing regime at low THz frequencies, before plasma waves become significant - allows the rectified current to be obtained from either the simulated or measured electrostatic transport properties of the MOSFET, as shown in \Figrefb{fig:Schematic+DC}. 
We find that the $\Re_I (U_{GS})$-curve obtained by our 2D hydrodynamic simulations at low frequency (0.5~THz) is in quantitative agreement with the analytical approximations of the current responsivity $\Re_I (U_{GS}) = \Delta I_{calc.}/P_S^{\rm{0.5 THz}}$, calculated with Eq.~(\ref{eq:Sako}) (see \Figrefc{fig:Sim}).} This represent a robust validation of our model and its numerical implementation for CMOS at low frequency and for small signals, i.e., for conditions, for which Eq.~(\ref{eq:Sako}) has proven abundantly in the literature to yield good agreement with experimental results. With regard to the frequency dependence of $\Re_I$, the strong decrease of the responsivity, which one observes in \Figrefc{fig:Sim} is consistent with  simulations performed for a simplified double-gate MOSFET in \cite{Linn2020}. In the experiments of \cite{Ludwig2024_Si_ADS}, a weaker frequency dependence was found, which may be attributed to the frequency dependence of the antenna impedance, which was neglected in the present work.

\textcolor{black}{In \Figrefd{fig:Sim}, we present additional simulations of the frequency-dependent current responsivity for a fixed gate voltage of $U_{GS}=0.5\,$V, providing clearer insight into the roll-off of the device's intrinsic detector response. To highlight the significance of plasmonic effects on the THz response of the 65-nm Si MOSFET, we compare two cases: one where the convective term ($j^2/n$) is included in the 2D hydrodynamic simulations (solid blue squares) and one where it is neglected (open blue squares). Consistent with other 1D hydrodynamic transport simulations of 65-nm Si MOSFETs reported in \cite{Ludwig2024_Si_ADS}, we observe that plasmonic effects are not dominant in these low-mobility devices, with enhancements of around 20\% at 1 THz (solid green squares). The simulations also reveal that the intrinsic detector responsivity follows an approximate $f^{-2}$ decay between 0.5 and 3 THz, before steepening to around $f^{-4}$ at higher THz frequencies. }

Finally, we present in \Figrefe{fig:Sim}) results of large-signal HDM simulations to showcase the stability of our numerical method. The time dependence of the drain current $I_D$ is presented in the lower panel of \Figrefe{fig:Sim}) for different THz amplitudes $U_a$. We find that the well-balanced HLLC approximate Riemann solver is numerically stable even in case of enormous source terms ($U_a = 1000\,$mV). The upper panel of \Figrefd{fig:Sim} depicts the periodic steady-state current responsivity as a function of $U_a$. At larger THz amplitudes, the rectified current $\Delta I$ saturates. 
This is a consequence of carrier velocity saturation in the high-field regime, which is accounted for by the Caughey-Thomas mobility model \cite{Caughey1967}. 

\section{Conclusion}
To summarize, we have successfully implemented and applied a numerical method for 2D hyperbolic partial differential equations with strong source terms - the well-balanced HLLC approximate Riemann solver. It is well suited for the hydrodynamic modeling of electronic components in the THz frequency range.  
We tested the solver by simulating the THz response of a 65-nm Si CMOS TeraFET detector. The well-balanced HLLC approximate Riemann solver is found to be numerically stable and positivity-preserving for the carrier density even in the presence of strong doping gradients and large external THz excitations.   
With the use of the 2D Poisson equation for the calculation of the DC and THz electric fields, the solver copes with situations beyond the gradual channel approximation valid for the ultra-high frequency simulations.
Our work extends recent advances in the field of plasma wave simulations in III-V compound semiconductors \cite{Bhardwaj2018, bhardwaj_electronicelectromagnetic_2019} or graphene \cite{cosme_tethys_2023} devices. It is expected that the implementation method presented here for the isothermal case and for bulk semiconductors can be readily extended to the full hydrodynamic transport equations (including carrier heating) for the simulation of highly nonlinear carrier transport phenomena in semiconductor devices arising in the THz frequency regime upon large-signal excitation.

\section*{Acknowledgment}
This work was funded by DFG RO 770/40-1, 770/40-2. We thank Prof. Dr.–Ing. C. Jungemann from the RWTH Aachen University for supplying us with his very helpful lecture notes on Numerical Device Simulation. 
\bibliography{THz}
\bibliographystyle{IEEEtran}

\end{document}

%% file: macros.tex
\newcommand{\Figref}[1]{Fig.~\ref{#1}}
\newcommand{\Figrefa}[1]{Fig.~\ref{#1}(a)}
\newcommand{\Figrefb}[1]{Fig.~\ref{#1}(b)}
\newcommand{\Figrefc}[1]{Fig.~\ref{#1}(c)}
\newcommand{\Figrefd}[1]{Fig.~\ref{#1}(d)}
\newcommand{\Figrefe}[1]{Fig.~\ref{#1}(e)}


%% file: main_IEEE.bbl
\begin{thebibliography}{10}
\providecommand{\url}[1]{#1}
\csname url@samestyle\endcsname
\providecommand{\newblock}{\relax}
\providecommand{\bibinfo}[2]{#2}
\providecommand{\BIBentrySTDinterwordspacing}{\spaceskip=0pt\relax}
\providecommand{\BIBentryALTinterwordstretchfactor}{4}
\providecommand{\BIBentryALTinterwordspacing}{\spaceskip=\fontdimen2\font plus
\BIBentryALTinterwordstretchfactor\fontdimen3\font minus \fontdimen4\font\relax}
\providecommand{\BIBforeignlanguage}[2]{{%
\expandafter\ifx\csname l@#1\endcsname\relax
\typeout{** WARNING: IEEEtran.bst: No hyphenation pattern has been}%
\typeout{** loaded for the language `#1'. Using the pattern for}%
\typeout{** the default language instead.}%
\else
\language=\csname l@#1\endcsname
\fi
#2}}
\providecommand{\BIBdecl}{\relax}
\BIBdecl

\bibitem{Dyakonov1993}
\BIBentryALTinterwordspacing
M.~Dyakonov and M.~Shur, ``{Shallow water analogy for a ballistic field effect transistor: New mechanism of plasma wave generation by dc current},'' \emph{Physical Review Letters}, vol.~71, no.~15, pp. 2465--2468, oct 1993. [Online]. Available: \url{http://link.aps.org/doi/10.1103/PhysRevLett.71.2465}
\BIBentrySTDinterwordspacing

\bibitem{Crowne1997}
F.~Crowne, ``Contact boundary conditions and the {Dyakonov–Shur} instability in high electron mobility transistors,'' \emph{Applied Physics Letters}, vol.~82, p. 1242–1254, 1997.

\bibitem{Mendl2018}
C.~Mendl and A.~Lucas, ``Dyakonov-{Shur} instability across the ballistic-to-hydrodynamic crossover,'' \emph{Applied Physics Letters}, vol. 112, p. 124101, 2020.

\bibitem{Kargar2018}
Z.~Kargar, T.~Linn, and C.~Jungemann, ``Investigation of the dyakonov–shur instability for thz wave generation based on the boltzmann transport equation,'' \emph{Semicond. Sci. Technol.}, vol.~33, p. 104001, 2018.

\bibitem{noei_numerical_2020}
\BIBentryALTinterwordspacing
M.~Noei, T.~Linn, and C.~Jungemann, ``A numerical approach to quasi-ballistic transport and plasma oscillations in junctionless nanowire transistors,'' vol.~19, no.~3, pp. 975--986. [Online]. Available: \url{https://link.springer.com/10.1007/s10825-020-01488-4}
\BIBentrySTDinterwordspacing

\bibitem{Dyakonov1996}
M.~Dyakonov and M.~Shur, ``{Detection, mixing, and frequency multiplication of terahertz radiation by two-dimensional electronic fluid},'' \emph{IEEE Transactions on Electron Devices}, vol.~43, no.~3, pp. 380–--387, 1996.

\bibitem{Boppel2012}
S.~Boppel, A.~Lisauskas, M.~Mundt, D.~Seliuta, L.~Minkevicius, I.~Kasalynas, G.~Valusis, M.~Mittendorff, S.~Winnerl, V.~Krozer, and H.~G. Roskos, ``{CMOS} integrated antenna-coupled field-effect transistors for the detection of radiation from 0.2 to 4.3~{THz},'' \emph{IEEE Transactions on Microwave Theory and Techniques}, vol.~60, no.~12, pp. 3834--3843, 2012.

\bibitem{Jungemann2022}
C.~Jungemann, M.~Noei, and T.~Linn, ``Device simulation of the {Dyakonov-Shur} plasma instability for {THz} wave generation,'' in \emph{{Proc. IEEE Latin American Electron Devices Conference (LAEDC)}}, 2022.

\bibitem{Knap2004}
W.~Knap, J.~Lusakowski, T.~Parenty, S.~Bollaert, A.~Cappy, V.~V. Popov, and M.~S. Shur, ``{Terahertz emission by plasma waves in 60 nm gate high electron mobility transistors},'' \emph{Applied Physics Letters}, vol.~84, no.~13, pp. 2331--2333, mar 2004.

\bibitem{Lisauskas2009}
\BIBentryALTinterwordspacing
A.~Lisauskas, U.~Pfeiffer, E.~{\"{O}}jefors, P.~H. Bol{\'{i}}var, D.~Glaab, and H.~G. Roskos, ``{Rational design of high-responsivity detectors of terahertz radiation based on distributed self-mixing in silicon field-effect transistors},'' \emph{Journal of Applied Physics}, vol. 105, no.~11, p. 114511, 2009. [Online]. Available: \url{https://dx.doi.org/10.1063/1.3140611}
\BIBentrySTDinterwordspacing

\bibitem{Lisauskas2013mix}
A.~Lisauskas, S.~Boppel, M.~Mundt, V.~Krozer, and H.~G. Roskos, ``Subharmonic mixing with field-effect transistors: Theory and experiment at 639~{GHz} high above f$_t$,'' \emph{IEEE SENSORS JOURNAL}, vol.~13, no.~1, p.~9, 2013.

\bibitem{Glaab2010}
\BIBentryALTinterwordspacing
D.~Glaab, S.~Boppel, A.~Lisauskas, U.~Pfeiffer, E.~{\"{O}}jefors, and H.~G. Roskos, ``{Terahertz heterodyne detection with silicon field-effect transistors},'' \emph{Applied Physics Letters}, vol.~96, no.~4, p. 042106, jan 2010. [Online]. Available: \url{http://aip.scitation.org/doi/10.1063/1.3292016}
\BIBentrySTDinterwordspacing

\bibitem{BoppelCMOSDetector}
S.~Boppel, A.~Lisauskas, A.~Max, V.~Krozer, and H.~G. Roskos, ``{CMOS detector arrays in a virtual 10-kilopixel camera for coherent terahertz real-time imaging},'' \emph{{Optics Letters}}, vol.~{37}, no.~{4}, pp. {536--538}, {FEB 15} {2012}.

\bibitem{Roadmap2021}
G.~Valušis, A.~Lisauskas, H.~Yuan, W.~Knap, and H.~G. Roskos, ``Roadmap of terahertz imaging 2021,'' \emph{Sensors}, vol.~21, p. 4092, 2021.

\bibitem{yuan2023a}
H.~Yuan, A.~Lisauskas, M.~D. Thomson, and H.~G. Roskos, ``600-{GHz} {Fourier} imaging based on heterodyne detection at the 2nd sub-harmonic,'' \emph{Optics Express}, vol.~31, no.~24, pp. 40\,856--40\,870, 2023.

\bibitem{wiecha_antenna-coupled_2021}
\BIBentryALTinterwordspacing
M.~M. Wiecha, R.~Kapoor, A.~V. Chernyadiev, K.~Ikamas, A.~Lisauskas, and H.~G. Roskos, ``Antenna-coupled field-effect transistors as detectors for terahertz near-field microscopy,'' \emph{Nanoscale Advances}, vol.~3, no.~6, pp. 1717--1724, 2021. [Online]. Available: \url{http://xlink.rsc.org/?DOI=D0NA00928H}
\BIBentrySTDinterwordspacing

\bibitem{hillger_terahertz_2019}
P.~Hillger, J.~Grzyb, R.~Jain, and U.~R. Pfeiffer, ``Terahertz imaging and sensing applications with silicon-based technologies,'' \emph{IEEE Transactions on Terahertz Science and Technology}, vol.~9, no.~1, pp. 1--19, Jan. 2019.

\bibitem{Ikamas2018}
K.~Ikamas, D.~Cibiraite, A.~Lisauskas, M.~Bauer, V.~Krozer, and H.~G. Roskos, ``Broadband terahertz power detectors based on 90-nm silicon {CMOS} transistors with flat responsivity up to 2.2~{THz},'' \emph{IEEE Electron Device Letters}, vol.~39, no.~9, pp. 1413--1416, sep 2018.

\bibitem{Hou2017}
H.~W. Hou, Z.~Liu, J.~H. Teng, T.~Palacios, and S.~J. Chua, ``High temperature terahertz detectors realized by a {GaN} high electron mobility transistor,'' \emph{Scientifc Reports}, vol.~7, p. 46664, 2017.

\bibitem{Regensburger2018_CW}
S.~Regensburger, A.~K. Mukherjee, S.~Sch{\"{o}}nhuber, M.~A. Kainz, S.~Winnerl, J.~M. Klopf, H.~Lu, A.~C. Gossard, K.~Unterrainer, and S.~Preu, ``Broadband terahertz detection with zero-bias field-effect transistors between 100~{GHz} and 11.8~{THz} with a noise equivalent power of 250~{pW}/$\sqrt{{Hz}}$ at 0.6~{THz},'' \emph{IEEE Transactions on Terahertz Science and Technology}, vol.~8, no.~4, pp. 465--471, 2018.

\bibitem{Bauer2019}
M.~Bauer, A.~R{\"{a}}mer, S.~A. Chevtchenko, K.~Y. Osipov, D.~{\v{C}}ibiraitė, S.~Pralgauskaitė, K.~Ikamas, A.~Lisauskas, W.~Heinrich, V.~Krozer, and H.~G. Roskos, ``A high-sensitivity {AlGaN}/{GaN} {HEMT} terahertz detector with integrated broadband bow-tie antenna,'' \emph{IEEE Transactions on Terahertz Science and Technology}, vol.~9, no.~4, pp. 430--444, 2019.

\bibitem{Regensburger2024}
S.~Regensburger, F.~Ludwig, S.~Winnerl, J.~M. Klopf, H.~Lu, H.~G. Roskos, and S.~Preu, ``Mapping the slow and fast photoresponse of field-effect transistors to terahertz and infrared radiation,'' \emph{Optics Express}, vol.~32, no.~5, p. 8458, 2024.

\bibitem{Zak2014}
A.~Zak, M.~A. Andersson, M.~Bauer, J.~Matukas, A.~Lisauskas, H.~G. Roskos, and J.~Stake, ``Antenna-integrated 0.6~{THz} {FET} direct detectors based on {CVD} graphene,'' \emph{Nano Letters}, vol.~14, no.~10, pp. 5834--5838, 2014.

\bibitem{Generalov2017}
A.~A. Generalov, M.~A. Andersson, X.~Yang, A.~Vorobiev, and J.~Stake, ``A heterodyne graphene {FET} detector at 400~{GHz},'' in \emph{2017 42nd International Conference on Infrared, Millimeter, and Terahertz Waves (IRMMW-THz)}, 2017, pp. 1--2.

\bibitem{Viti2020}
L.~Viti, D.~G. Purdie, A.~Lombardo, A.~C. Ferrari, and M.~S. Vitiello, ``{HBN}-encapsulated, graphene-based, room-temperature terahertz receivers, with high speed and low noise,'' \emph{Nano Letters}, vol.~20, p. 3169–77, 2020.

\bibitem{Generalov2024}
F.~Ludwig, A.~Generalov, J.~Holstein, M.~A., K.~Viisanen, M.~Prunilla, and H.~G. Roskos, ``Terahertz detection with graphene {FETs}: Photothermoelectric and resistive self-mixing contributions to the detector response,'' \emph{ACS Applied Electronic Materials}, vol.~6, pp. 2197--2212, 2024.

\bibitem{Viti2021}
L.~Viti, A.~R. Cadore, X.~Yang, A.~Vorobiev, J.~E. Muench, K.~Watanabe, T.~Taniguchi, J.~Stake, A.~C. Ferrari, and M.~S. Vitiello, ``Thermoelectric graphene photodetectors with sub-nanosecond response times at terahertz frequencies,'' \emph{Nanophotonics}, vol.~10, no.~1, pp. 89--98, 2021.

\bibitem{Mateos2020}
P.~Martín-Mateos, D.~Čibiraitė Lukenskienė, R.~Barreiro, C.~{de Dios}, A.~Lisauskas, V.~Krozer, and P.~Acedo, ``Hyperspectral terahertz imaging with electro-optic dual combs and a {FET}-based detector,'' \emph{Scientific Reports}, vol.~10, p. 14429, 2020.

\bibitem{Drexler2012}
C.~Drexler, N.~Dyakonova, P.~Olbrich, J.~Karch, M.~Schafberger, K.~Karpierz, Y.~Mityagin, M.~B. Lifshits, F.~Teppe, O.~Klimenko, Y.~M. Meziani, W.~Knap, and S.~D. Ganichev, ``{Helicity sensitive terahertz radiation detection by field effect transistors},'' \emph{Journal of Applied Physics}, vol. 111, no.~12, p. 124504, 2012.

\bibitem{Bandurin2018a}
D.~A. Bandurin, D.~Svintsov, I.~Gayduchenko, S.~G. Xu, A.~Principi, M.~Moskotin, I.~Tretyakov, and et~al., ``Resonant terahertz detection using graphene plasmons,'' \emph{Nature Communications}, vol.~9, p. 5392, 2018.

\bibitem{Delgado-Notario2024}
J.~M. Caridad, O.~Castelló, S.~M.~L. Baptista, T.~Taniguchi, K.~Watanabe, H.~G. Roskos, and J.~A. Delgado-Notario, ``Room-temperature plasmon-assisted resonant {THz} detection in single-layer graphene transistors,'' \emph{Nano Letters}, vol.~24, p. 935–942, 2024.

\bibitem{Soltani2020}
A.~Soltani, F.~Kuschewski, M.~Bonmann, A.~Generalov, A.~Vorobiev, F.~Ludwig, M.~M. Wiecha, D.~Čibiraitė, F.~Walla, S.~Winnerl, S.~C. Kehr, L.~M. Eng, J.~Stake, and H.~G. Roskos, ``Direct nanoscopic observation of plasma waves in the channel of a graphene field-effect transistor,'' \emph{Light: Science and Applications}, vol.~9, p.~97, 2020.

\bibitem{Ludwig2019}
F.~Ludwig, M.~Bauer, A.~Lisauskas, and H.~G. Roskos, ``Circuit-based hydrodynamic modeling of {AlGaN}/{GaN} {HEMTs},'' in \emph{ESSDERC 2019 - 49th European Solid-State Device Research Conference (ESSDERC)}, 2019, pp. 270--273.

\bibitem{Ludwig2024_Si_ADS}
F.~Ludwig, J.~Holstein, A.~Krysl, A.~Lisauskas, and H.~G. Roskos, ``Modeling of antenna-coupled {Si} {MOSFETs} in the terahertz frequency range,'' \emph{IEEE Transactions on Terahertz Science and Technology}, vol.~14, no.~3, pp. 1--11, 2024.

\bibitem{Stake2016}
M.~A. Andersson and J.~Stake, ``An accurate empirical model based on {Volterra} series for {FET} power detectors,'' \emph{IEEE Transactions on Microwave Theory and Techniques}, vol.~64, no.~5, pp. 1431--1441, may 2016.

\bibitem{Millithaler09}
\BIBentryALTinterwordspacing
J.-F. Millithaler, J.~Pousset, L.~Reggiani, P.~Ziade, H.~Marinchio, L.~Varani, C.~Palermo, J.~Mateos, T.~González, S.~Perez, and D.~Pardo, ``Monte {Carlo} investigation of terahertz plasma oscillations in gated ultrathin channel of n-{InGaAs},'' \emph{Applied Physics Letters}, vol.~95, no.~15, p. 152102, Oct. 2009. [Online]. Available: \url{https://doi.org/10.1063/1.3248096}
\BIBentrySTDinterwordspacing

\bibitem{Millithaler11}
J.-F. Millithaler, J.~Pousset, L.~Reggiani, H.~Marinchio, L.~Varani, C.~Palermo, P.~Ziade, J.~Mateos, T.~González, and S.~Perez, ``Transconductance characteristics and plasma oscillations in nanometric ingaas field effect transistors,'' \emph{Solid-State Electronics}, vol.~56, no.~1, pp. 116--119, 2011.

\bibitem{Mateos12}
J.~Mateos and T.~Gonzalez, ``Plasma enhanced terahertz rectification and noise in ingaas hemts,'' \emph{IEEE Transactions on Terahertz Science and Technology}, vol.~2, no.~5, pp. 562--569, 2012.

\bibitem{oberman2011monte}
A.~Oberman and Y.~Ruan, ``Monte carlo methods for high-dimensional elliptic partial differential equations,'' \emph{Multiscale Modeling \& Simulation}, vol.~9, no.~2, pp. 790--812, 2011.

\bibitem{dimarco2018uncertainty}
G.~Dimarco, L.~Pareschi, and M.~Zanella, ``Uncertainty quantification for kinetic equations: A comparison of stochastic methods,'' \emph{Journal of Computational Physics}, vol. 371, pp. 397--421, 2018.

\bibitem{Blotekjaer1970}
K.~Bløtekjær, ``Transport equations for electrons in two-valley semiconductors,'' \emph{IEEE Transactions on Electron Devices}, vol.~17, p. 38–47, 1970.

\bibitem{grasser2003}
T.~Grasser, T.-W. Tang, H.~Kosina, and S.~Selberherr, ``{A review of hydrodynamic and energy-transport models for semiconductor device simulation},'' \emph{Proceedings of the IEEE}, vol.~91, no.~2, pp. 251--274, 2003.

\bibitem{Gutin2012}
A.~Gutin, V.~Kachorovskoii, and A.~M.~M. Shur, ``Plasmonic terahertz detector response to high intensities,'' \emph{Journal of Applied Physics}, vol. 112, p. 014508, 2012.

\bibitem{Rudin2014}
S.~Rudin, G.~Rupper, A.~Gutin, and M.~Shur, ``Theory and measurement of plasmonic terahertz detector response to large signals,'' \emph{Journal of Applied Physics}, vol. 115, p. 064503, 2014.

\bibitem{Scharfetter1969}
D.~Scharfetter and H.~Gummel, ``Large-signal analysis of a silicon read diode oscillator,'' \emph{IEEE Transactions on Electron Devices}, vol.~16, no.~1, pp. 64--77, 1969.

\bibitem{Selberherr1984}
S.~Selberherr, \emph{Analysis and Simulation of Semiconductor Devices}, 1st~ed.\hskip 1em plus 0.5em minus 0.4em\relax Springer Vienna, 1984.

\bibitem{Liu2019}
X.~Liu and M.~Shur, ``An efficient {TCAD} model for {TeraFET} detectors,'' \emph{IEEE Radio and Wireless Symposium}, p. 1–4, 2019.

\bibitem{Linn2020}
T.~Linn, K.~Bittner, Brachtendorf, H.G., and C.~Jungemann, ``Simulation of {THz} oscillations in semiconductor devices based on balance equations,'' \emph{J Sci Comput}, vol.~85, p.~6, 2020.

\bibitem{Leveque2007}
\BIBentryALTinterwordspacing
R.~J. LeVeque, \emph{\BIBforeignlanguage{en}{Finite {Difference} {Methods} for {Ordinary} and {Partial} {Differential} {Equations}: {Steady}-{State} and {Time}-{Dependent} {Problems}}}.\hskip 1em plus 0.5em minus 0.4em\relax Society for Industrial and Applied Mathematics, Jan. 2007. [Online]. Available: \url{http://epubs.siam.org/doi/book/10.1137/1.9780898717839}
\BIBentrySTDinterwordspacing

\bibitem{Bhardwaj2018}
S.~Bhardwaj, F.~L. Teixeira, and J.~L. Volakis, ``Fast modeling of terahertz plasma-wave devices using unconditionally stable fdtd methods,'' \emph{IEEE Journal on Multiscale and Multiphysics Computational Techniques}, vol.~3, pp. 29--36, 2018.

\bibitem{bhardwaj_electronicelectromagnetic_2019}
S.~Bhardwaj, ``\BIBforeignlanguage{en}{Electronic–electromagnetic multiphysics modeling for terahertz plasmonics: {A} review},'' \emph{\BIBforeignlanguage{en}{IEEE Journal on Multiscale and Multiphysics Computational Techniques}}, vol.~4, pp. 307--316, 2019.

\bibitem{cosme_tethys_2023}
P.~Cosme, J.~S. Santos, J.~P.~S. Bizarro, and I.~Figueiredo, ``{TETHYS}: {A} simulation tool for graphene hydrodynamic models,'' \emph{Computer Physics Communications}, vol. 282, p. 108550, 2023.

\bibitem{kurganov_second-order_2007}
A.~Kurganov and G.~Petrova, ``\BIBforeignlanguage{en}{A second-order well-balanced positivity preserving central-upwind scheme for the {Saint}-{Venant} system},'' \emph{\BIBforeignlanguage{en}{Communications in Mathematical Sciences}}, vol.~5, no.~1, pp. 133--160, 2007.

\bibitem{kurganov_finite-volume_2018}
A.~Kurganov, ``\BIBforeignlanguage{en}{Finite-volume schemes for shallow-water equations},'' \emph{\BIBforeignlanguage{en}{Acta Numerica}}, vol.~27, pp. 289--351, May 2018.

\bibitem{Cravero2018}
I.~Cravero, G.~Puppo, M.~Semplice, and G.~Visconti, ``Cool {WENO} schemes,'' \emph{Computers \& Fluids}, vol. 169, pp. 71--86, 2018.

\bibitem{Cravero2019}
I.~Cravero, M.~Semplice, and G.~Visconti, ``Optimal definition of the nonlinear weights in multidimensional central {WENOZ} reconstructions,'' \emph{{SIAM} Journal on Numerical Analysis}, vol.~57, no.~5, pp. 2328--2358, 2019.

\bibitem{Toro1994}
\BIBentryALTinterwordspacing
E.~F. Toro, M.~Spruce, and W.~Speares, ``\BIBforeignlanguage{en}{Restoration of the contact surface in the {HLL}-{Riemann} solver},'' \emph{\BIBforeignlanguage{en}{Shock Waves}}, vol.~4, no.~1, pp. 25--34, Jul. 1994. [Online]. Available: \url{http://link.springer.com/10.1007/BF01414629}
\BIBentrySTDinterwordspacing

\bibitem{Toro2009}
E.~F. Toro, \emph{Riemann solvers and numerical methods for fluid dynamics: a practical introduction}, 3rd~ed.\hskip 1em plus 0.5em minus 0.4em\relax USA: Dordrecht New York: Springer, 2009.

\bibitem{Audusse2015}
\BIBentryALTinterwordspacing
E.~Audusse, C.~Chalons, and P.~Ung, ``A simple well-balanced and positive numerical scheme for the shallow-water system,'' \emph{Communications in Mathematical Sciences}, vol.~13, no.~5, pp. 1317--1332, 2015. [Online]. Available: \url{http://www.intlpress.com/site/pub/pages/journals/items/cms/content/vols/0013/0005/a011/}
\BIBentrySTDinterwordspacing

\bibitem{Cole1989}
D.~Cole and J.~Johnson, ``Accounting for incomplete ionization in modeling silicon based semiconductor devices,'' in \emph{Proceedings of the Workshop on Low Temperature Semiconductor Electronics,}, 1989, pp. 73--77.

\bibitem{HandbookOptoSim}
J.~Piprek, \emph{Handbook of Optoelectronic Device Modeling and Simulation}, 1st~ed.\hskip 1em plus 0.5em minus 0.4em\relax CRC Press Taylor and Francis Group, 2017, vol.~2, pp. 733--771.

\bibitem{Caughey1967}
D.~Caughey and R.~Thomas, ``Carrier mobilities in silicon empirically related to doping and field,'' \emph{Proceedings of the IEEE}, vol.~55, no.~12, pp. 2192--2193, 1967.

\bibitem{VanRoosbroeck1950}
\BIBentryALTinterwordspacing
W.~Van~Roosbroeck, ``\BIBforeignlanguage{en}{Theory of the {Flow} of {Electrons} and {Holes} in {Germanium} and {Other} {Semiconductors}},'' \emph{\BIBforeignlanguage{en}{Bell System Technical Journal}}, vol.~29, no.~4, pp. 560--607, Oct. 1950. [Online]. Available: \url{https://ieeexplore.ieee.org/document/6772705}
\BIBentrySTDinterwordspacing

\bibitem{Lombardi1988}
C.~Lombardi, S.~Manzini, A.~Saporito, and M.~Vanzi, ``A physically based mobility model for numerical simulation of nonplanar devices,'' \emph{IEEE Transactions on Computer-Aided Design of Integrated Circuits and Systems}, vol.~7, no.~11, pp. 1164--1171, 1988.

\bibitem{Masetti1983}
G.~Masetti, M.~Severi, and S.~Solmi, ``Modeling of carrier mobility against carrier concentration in arsenic-, phosphorus-, and boron-doped silicon,'' \emph{IEEE Transactions on Electron Devices}, vol.~30, no.~7, pp. 764--769, 1983.

\bibitem{Shockley1952}
\BIBentryALTinterwordspacing
W.~Shockley and W.~T. Read, ``Statistics of the recombinations of holes and electrons,'' \emph{Phys. Rev.}, vol.~87, pp. 835--842, Sep 1952. [Online]. Available: \url{https://link.aps.org/doi/10.1103/PhysRev.87.835}
\BIBentrySTDinterwordspacing

\bibitem{AufderMaur2008}
\BIBentryALTinterwordspacing
M.~A. der Maur, ``A multiscale simulation environment for electronic and optoelectronic devices,'' dissertation, Università degli Studi di Roma – Tor Vergata, 2008. [Online]. Available: \url{http://www.optolab.uniroma2.it/images/stories/team/aufdermaur/thesis_aufdermaur.pdf}
\BIBentrySTDinterwordspacing

\bibitem{vasileska_computational_2017}
\BIBentryALTinterwordspacing
D.~Vasileska, S.~M.~Goodnick, and G.~Klimeck, \emph{Computational Electronics: Semiclassical and Quantum Device Modeling and Simulation}, 1st~ed.\hskip 1em plus 0.5em minus 0.4em\relax {CRC} Press. [Online]. Available: \url{https://www.taylorfrancis.com/books/9781420064841}
\BIBentrySTDinterwordspacing

\bibitem{NumericalRecipes2007}
W.~H. Press, S.~A. Teukolsky, W.~T. Vetterling, and B.~P. Flannery, \emph{Numerical Recipes 3rd Edition: The Art of Scientific Computing}, 3rd~ed.\hskip 1em plus 0.5em minus 0.4em\relax USA: Cambridge University Press, 2007.

\bibitem{Saad2010}
I.~Saad, M.~L.~P. Tan, M.~T. Ahmadi, R.~Ismail, and V.~K. Arora, ``The dependence of saturation velocity on temperature, inversion charge and electric field in a nanoscale {MOSFET},'' \emph{International Journal of Nanoelectronics and Materials}, vol.~3, pp. 17--34, 2010.

\bibitem{green_intrinsic_1990}
M.~A. Green, ``\BIBforeignlanguage{en}{Intrinsic concentration, effective densities of states, and effective mass in silicon},'' \emph{\BIBforeignlanguage{en}{Journal of Applied Physics}}, vol.~67, no.~6, pp. 2944--2954, Mar. 1990.

\bibitem{Fung2004}
S.~Fung, H.~Huang, S.~Cheng, K.~Cheng, S.~Wang, Y.~Wang, Y.~Yao, C.~Chu, S.~Yang, W.~Liang, Y.~Leung, C.~Wu, C.~Lin, S.~Chang, S.~Wu, C.~Nieh, C.~Chen, T.~Lee, Y.~Jin, S.~Chen, L.~Lin, Y.~Chiu, H.~Tao, C.~Fu, S.~Jang, K.~Yu, C.~Wang, T.~Ong, Y.~See, C.~Diaz, M.~Liang, and Y.~Sun, ``65nm {CMOS} high speed, general purpose and low power transistor technology for high volume foundry application,'' in \emph{Digest of Technical Papers. 2004 Symposium on VLSI Technology, 2004.}, 2004, pp. 92--93.

\bibitem{Yang2011}
\BIBentryALTinterwordspacing
S.~Yang, J.~Sheu, M.~Ieong, M.~Chiang, T.~Yamamoto, J.~Liaw, S.~Chang, Y.~Lin, T.~Hsu, J.~Hwang, J.~Ting, C.~Wu, K.~Ting, F.~Yang, C.~Liu, I.~Wu, Y.~Chen, S.~Chent, K.~Chen, J.~Cheng, M.~Tsai, W.~Chang, R.~Chen, C.~Chen, T.~Lee, C.~Lin, S.~Yang, Y.~Sheu, J.~Tzeng, L.~Lu, S.~Jang, C.~Diaz, and Y.~Mii, ``\BIBforeignlanguage{en}{28nm metal-gate high-{K} {CMOS} {SoC} technology for high-performance mobile applications},'' in \emph{\BIBforeignlanguage{en}{2011 {IEEE} {Custom} {Integrated} {Circuits} {Conference} ({CICC})}}.\hskip 1em plus 0.5em minus 0.4em\relax San Jose, CA, USA: IEEE, Sep. 2011, pp. 1--5. [Online]. Available: \url{http://ieeexplore.ieee.org/document/6055355/}
\BIBentrySTDinterwordspacing

\bibitem{Singanamalla2007}
\BIBentryALTinterwordspacing
R.~Singanamalla, H.~Y. Yu, B.~Van~Daele, S.~Kubicek, and K.~De~Meyer, ``\BIBforeignlanguage{en}{Effective {Work}-{Function} {Modulation} by {Aluminum}-{Ion} {Implantation} for {Metal}-{Gate} {Technology} \$({\textbackslash}hbox\{{Poly}-{Si}/{TiN}/{SiO}\}\_\{2\})\$},'' \emph{\BIBforeignlanguage{en}{IEEE Electron Device Letters}}, vol.~28, no.~12, pp. 1089--1091, Dec. 2007. [Online]. Available: \url{http://ieeexplore.ieee.org/document/4383548/}
\BIBentrySTDinterwordspacing

\bibitem{Kadoshima2009}
\BIBentryALTinterwordspacing
M.~Kadoshima, T.~Matsuki, S.~Miyazaki, K.~Shiraishi, T.~Chikyo, K.~Yamada, T.~Aoyama, Y.~Nara, and Y.~Ohji, ``\BIBforeignlanguage{en}{Effective-{Work}-{Function} {Control} by {Varying} the {TiN} {Thickness} in {Poly}-{Si}/{TiN} {Gate} {Electrodes} for {Scaled} {High}- \$k\$ {CMOSFETs}},'' \emph{\BIBforeignlanguage{en}{IEEE Electron Device Letters}}, vol.~30, no.~5, pp. 466--468, May 2009. [Online]. Available: \url{https://ieeexplore.ieee.org/document/4808212}
\BIBentrySTDinterwordspacing

\bibitem{Robertson2009}
\BIBentryALTinterwordspacing
J.~Robertson, ``\BIBforeignlanguage{en}{Band offsets and work function control in field effect transistors},'' \emph{\BIBforeignlanguage{en}{Journal of Vacuum Science \& Technology B: Microelectronics and Nanometer Structures Processing, Measurement, and Phenomena}}, vol.~27, no.~1, pp. 277--285, Jan. 2009. [Online]. Available: \url{https://pubs.aip.org/jvb/article/27/1/277/1046345/Band-offsets-and-work-function-control-in-field}
\BIBentrySTDinterwordspacing

\bibitem{Jungemann2016}
C.~Jungemann, K.~Bittner, and H.~G. Brachtendorf, ``Simulation of plasma resonances in mosfets for thz-signal detection,'' \emph{Joint International EUROSOI Workshop and International Conference on Ultimate Integration on Silicon}, p. 48–51, 2016.

\bibitem{Ramo1939}
S.~Ramo, ``Currents induced by electron motion,'' \emph{Proceedings of the IRE}, vol.~27, no.~9, pp. 584--585, 1939.

\bibitem{Kim1991}
\BIBentryALTinterwordspacing
H.~Kim, H.~Min, T.~Tang, and Y.~Park, ``\BIBforeignlanguage{en}{An extended proof of the {Ramo}-{Shockley} theorem},'' \emph{\BIBforeignlanguage{en}{Solid-State Electronics}}, vol.~34, no.~11, pp. 1251--1253, Nov. 1991. [Online]. Available: \url{https://linkinghub.elsevier.com/retrieve/pii/0038110191900657}
\BIBentrySTDinterwordspacing

\bibitem{Yanenko1971}
\BIBentryALTinterwordspacing
N.~N. Yanenko, \emph{\BIBforeignlanguage{en}{The {Method} of {Fractional} {Steps}}}, M.~Holt, Ed.\hskip 1em plus 0.5em minus 0.4em\relax Berlin, Heidelberg: Springer Berlin Heidelberg, 1971. [Online]. Available: \url{http://link.springer.com/10.1007/978-3-642-65108-3}
\BIBentrySTDinterwordspacing

\bibitem{LeVeque2011}
\BIBentryALTinterwordspacing
R.~J. LeVeque, ``\BIBforeignlanguage{en}{A {Well}-{Balanced} {Path}-{Integral} f-{Wave} {Method} for {Hyperbolic} {Problems} with {Source} {Terms}},'' \emph{\BIBforeignlanguage{en}{Journal of Scientific Computing}}, vol.~48, no. 1-3, pp. 209--226, Jul. 2011. [Online]. Available: \url{http://link.springer.com/10.1007/s10915-010-9411-0}
\BIBentrySTDinterwordspacing

\bibitem{Harten1983}
\BIBentryALTinterwordspacing
A.~Harten, P.~D. Lax, and B.~V. Leer, ``\BIBforeignlanguage{en}{On {Upstream} {Differencing} and {Godunov}-{Type} {Schemes} for {Hyperbolic} {Conservation} {Laws}},'' \emph{\BIBforeignlanguage{en}{SIAM Review}}, vol.~25, no.~1, pp. 35--61, Jan. 1983. [Online]. Available: \url{http://epubs.siam.org/doi/10.1137/1025002}
\BIBentrySTDinterwordspacing

\bibitem{Bouchut2005}
\BIBentryALTinterwordspacing
F.~Bouchut, ``\BIBforeignlanguage{en}{An introduction to finite volume methods forhyperbolic conservation laws},'' \emph{\BIBforeignlanguage{en}{ESAIM: Proceedings}}, vol.~15, pp. 1--17, 2005. [Online]. Available: \url{http://www.esaim-proc.org/10.1051/proc:2005020}
\BIBentrySTDinterwordspacing

\bibitem{Oeje09}
E.~\"{O}jefors, U.~R. Pfeiffer, A.~Lisauskas, and H.~G. Roskos, ``A 0.65~{THz} focal-plane array in a quarter-micron {CMOS} process technology,'' \emph{IEEE Journal of Solid-State Circuits}, vol.~44, no.~7, pp. 1968--1976, 2009.

\bibitem{Sakowicz2011}
\BIBentryALTinterwordspacing
M.~Sakowicz, M.~B. Lifshits, O.~A. Klimenko, F.~Schuster, D.~Coquillat, F.~Teppe, and W.~Knap, ``{Terahertz responsivity of field effect transistors versus their static channel conductivity and loading effects},'' \emph{Journal of Applied Physics}, vol. 110, no.~5, p. 054512, sep 2011. [Online]. Available: \url{http://doi.org/10.1063/1.3632058}
\BIBentrySTDinterwordspacing

\end{thebibliography}
